\documentclass[%
showpacs,preprintnumbers,
amsmath,amssymb,
aps,
prb,
 reprint,
]{revtex4-2}

\usepackage{graphicx}
\usepackage{dcolumn}
\usepackage{float}
\usepackage{bm}

\usepackage[pagebackref=false,colorlinks,linkcolor=blue,filecolor=green,urlcolor=blue ,citecolor=blue]{hyperref}
\usepackage{mathdots}
\usepackage[normalem]{ulem}
\usepackage{comment}

\usepackage[T1]{fontenc}

\usepackage{xcolor}

\def\vahid{\textcolor{red}}

\usepackage{cancel}

\begin{document}

\title{$\mathcal{PT}$ and anti-$\mathcal{PT}$ phase transitions in a trimerized Su--Schrieffer--Heeger chain with nonreciprocal Rashba spin-orbit coupling
} 

\author{Milad Jangjan$^1$, Linhu Li$^2$, Longwen Zhou$^{3,4,5}$, and Mir Vahid Hosseini$^6$}
 \email[Corresponding author: ]{mv.hosseini@znu.ac.ir}
\affiliation{$^1$Institute of Physics, University of Rostock, Albert-Einstein-Stra{\ss}e 23-24, 
D-18059 Rostock, Germany} 
\affiliation{$^2$Quantum Science Center of Guangdong-Hong Kong-Macao Greater Bay Area (Guangdong), Shenzhen 518045, China}
\affiliation{$^3$College of Physics and Optoelectronic Engineering, Ocean University of China, Qingdao 266100, China}
\affiliation{$^4$Qingdao Key Laboratory of Advanced Optoelectronics, Qingdao 266100, China}
\affiliation{$^5$Engineering Research Center of Advanced Marine Physical Instruments and Equipment of MOE, Qingdao 266100, China}
\affiliation{$^6$Department of Physics, Faculty of Science, University of Zanjan, Zanjan 45371-38791, Iran}

\begin{abstract}
 We theoretically investigate a one-dimensional trimerized Su--Schrieffer--Heeger chain with three sublattices per unit cell subjected to a nonreciprocal Rashba spin-orbit coupling. Invoking a spin-flip symmetry, the non-Hermitian Hamiltonian decomposes into two independent spin sectors, enabling a spin-resolved analysis of non-Hermitian skin effects and system symmetries. We identify a rich phase diagram consisting of four bulk phases and two edge-state phases. The bulk phases include fully $\mathcal{PT}$-unbroken (real-spectrum) and fully anti-$\mathcal{PT}$-unbroken (imaginary-spectrum) regimes, as well as two mixed phases where one band remains on the real or imaginary axis while the other two form complex-conjugate pairs. The two edge-state phases correspond to topological edge modes with either $\mathcal{PT}$-unbroken (real) or anti-$\mathcal{PT}$-unbroken (imaginary) energies. Using non-Bloch band theory and Cardano's method, we derive closed-form expressions for phase boundaries and establish the bulk-edge correspondence for each spin sector. Calculations of Berry phase and directional inverse participation ratios confirm our analytical predictions. Our results provide a minimal platform for realizing spin-resolved non-Hermitian topology and edge-selective symmetry preservation, in which the bulk and edge states belong to distinct symmetry classes. 
\end{abstract}

\maketitle

\section{Introduction}\label{s1}

Non-Hermitian quantum physics \cite{NonHerQuanPhys} has recently attracted a lot of attention from both theoretical and experimental points of view \cite{NonHerPhys,NonHerRev1,NonHerRev2,NonHerRev3,NonHerRev4}. In traditional quantum mechanics, the Hermiticity of Hamiltonians ensures real eigenvalues. However, non-Hermitian Hamiltonians can still exhibit entirely real spectra when protected by symmetries such as parity–time ($\mathcal{PT}$) symmetry \cite{Pt1,Pt2} or more general pseudo-Hermitian structures \cite{PH1,PH2,PH3,dirIPR1}. In recent decades, non-Hermitian physics has provided a powerful framework, with a wide range of non-Hermitian Hamiltonians for describing open quantum systems \cite{QS1,QS2,QS3,QS4,QS6,QS7,PRACheng}, nonreciprocal transport phenomena~\cite{Non-rec1,Non-rec2,Non-rec3,Non-rec4}, and novel topological phases \cite{Non_Her_topo1,Non_Her_topo3,Non_Her_topo5,Non_Her_topo6,Non_Her_topo7}. 

One of the most exotic features of non-Hermiticity is the non-Hermitian skin effect (NHSE), in which an extensive number of bulk states accumulate at a boundary under open boundary conditions \cite{TopoSkin, NHSE1,NHSE2,NHSE3,NHSE5,NHSE7,NHSE8,NHSE9}. This effect traces back to the late 1990s when Hatano and Nelson \cite{Hatano1} introduced a one-dimensional disordered tight-binding lattice with nearest-neighbor nonreciprocal hopping. They revealed that non-Hermiticity, induced by nonreciprocity, can inhibit Anderson localization, establishing a mobility region that facilitates unidirectional transport \cite{Hatano1,Hatano2}. Recently, the NHSE has been explored extensively in both theoretical \cite{Theory1,Theory2,Theory3} and experimental \cite{Experiment1,Experiment2,Experiment3,Experiment4,Experiment5,Experiment6} platforms. The interplay between skin modes and criticality has further revealed the critical NHSE, where eigenstates discontinuously jump between distinct skin solutions at the critical point separating phases with different localization lengths \cite{LiCritical2020}.

The NHSE can also be interpreted as a topological phenomenon characterized by point-gap topology  \cite{pointgaptopo1,pointgaptopo2,pointgaptopo3,NonHerRev3}, invalidating the conventional bulk–boundary correspondence. As such, unlike the usual Bloch band theory that characterizes Hermitian systems, in non-Hermitian cases, the band theory has been generalized to the non-Bloch form \cite{NonBloch1,NHSE1,NHSE2,NonBloch4,NonBloch5,NHSE5}. It extends the concept of Brillouin zone (BZ) to the generalized Brillouin zone (GBZ) \cite{NHSE7,GBZ2,GBZ3,GBZ4,GBZ5}, with geometric characterizations through singularity structures in pseudospin vector space providing complementary insights \cite{LiGeometric2019,mandal2024identifying}. These developments correctly describe spectra and wavefunctions under open boundaries \cite{NonBloch1,NHSE3,NHSE1,NHSE8}. Thus,
incorporating the physics of NHSE is essential to establish the proper non-Hermitian bulk-boundary correspondence \cite{NHSE3,BBC2,BBC3,BBC4,BBC5,BBC7} in topological systems. 

One of the simplest models for studying topological phases in one dimension is the Su--Schrieffe--Heeger (SSH) model \cite{Su1979}. It was originally introduced to simulate polyacetylene consisting of a dimerized chain with alternating hopping amplitudes. The SSH chain supports two distinct insulating phases characterized by different winding numbers of the Bloch Hamiltonian \cite{Ryu2002,Asboth2016}. In its topologically nontrivial phase, the SSH model hosts zero-energy edge states localized at the chain boundaries, which are protected by chiral symmetry \cite{Heeger1988}. The SSH model has inspired extensive generalizations \cite{Li2014}, including systems with multiple sublattices (SSH$_n$) \cite{Jangjan2020,Jangjan2021,Jangjan2022,SSH3,PRBVerma,aghtouman2024dimerized} and spinful degrees of freedom \cite{SSHspin1,HermiSOC2016}. The experimental verification of topological edge states in SSH-type systems has been achieved across diverse platforms, including photonic waveguide arrays \cite{Blanco2016}, ultracold atoms in optical lattices \cite{Atala2013}, mechanical metamaterials \cite{Huber2016}, and electrical circuits enriched by $\mathcal{PT}$ and anti-$\mathcal{PT}$ symmetry \cite{PRLThomale}.  

The extension of SSH model to non-Hermitian cases has revealed a broad range of unique physics.
Unlike generic non-Hermitian lattices, the dimerized structure of SSH model allows for competing effects between skin localization versus topological edge protection. In non-Hermitian SSH model, the nonreciprocal hopping \cite{Lai2025} between dimers can drive the NHSE while simultaneously modifying the topological winding number in nontrivial ways \cite{Zeng2022,Liu2022}. $\mathcal{PT}$ symmetric SSH model reveals conjugated-pseudo-Hermiticity preserving bulk-boundary correspondence \cite{Halder2023Properties} and leading to a quantized complex Berry phase that serves as a topological invariant \cite{Lieu2018}. 
On the other hand,
anti-$\mathcal{PT}$ ($\mathcal{APT}$) symmetry, which protects imaginary spectra instead of real ones, can also support topological phases, inducing phase transitions characterized by non-Hermitian winding numbers~\cite{WuAPT2021, wu2025topological,Jangjan2024}. Recent theoretical advances have extended these concepts to long-range hopping SSH chains where gain and loss induce novel topological phase transitions~\cite{Zhang2024LongRange}, while exceptional points arising from hopping amplitude gradients provide additional topological signatures~\cite{Simon2024ExceptionalPoints}. Extensions to topological switches for the NHSE in cold-atom systems~\cite{LiSwitch2020}, $n$-root topological phases in photonic ring systems~\cite{Viedma2024nRoot}, size-dependent non-Hermitian effects~\cite{Siu2025SizeDependent}, and tunable NHSE configurations~\cite{Zhu2023TunableNHSE} have substantially broadened the theoretical landscape. Beyond these developments,
spin-orbit coupling provides another important avenue for manipulating Hermitian topological phases ~\cite{SSHspin1,SSH3,HermiSOC2016,HermiSOC2022,HermiSOC2020}. However, its interplay with non-Hermiticity remains less explored~\cite{Liu2022}, particularly in systems with multiple sublattices.

In this work, we study a trimerized SSH lattice with three sublattices per unit cell (SSH$_3$) in the presence of nonreciprocal Rashba spin-orbit coupling (RSOC). Exploiting a spin-flip symmetry, the full Hamiltonian analytically decomposes into two decoupled subsystems. This decomposition allows us to uncover spin-sector asymmetry, driving spin-dependent NHSEs and spin-resolved $\mathcal{PT}$/$\mathcal{APT}$ symmetry-breaking phase transitions. We further present two distinct GBZs for each subsystem, restoring the non-Hermitian bulk-boundary correspondence. We also derive analytical expressions for phase boundaries (using Cardano’s method), and localized edge states. These analytical results are supplemented by numerical calculations of Berry phases and directional inverse participation ratios to build the phase diagram.  We highlight a remarkable feature in which  $\mathcal{PT}$ and $\mathcal{APT}$ symmetries are separately preserved for bulk and edge states, so that the bulk spectrum remains real while the edge spectrum becomes imaginary, which we term \emph{bulk-edge symmetry class separation}. These edge modes are protected by fundamental symmetries, remaining strictly pinned at zero energy so long as these symmetries are preserved. 

The rest of the paper is organized as follows. In Sec.~\ref{s2}, we introduce our model and show that a spin-flip symmetry allows the Hamiltonian to be decomposed into two decoupled spin-resolved subsystems. In Sec.~\ref{s3}, we investigate symmetries of the system under open boundary conditions (OBCs). Then, the GBZ formulation is used under periodic boundary conditions (PBCs) to classify bulk spectral phases. In addition, we apply the similarity transformation under OBCs to reveal the properties of edge states associated to the corresponding symmetries. Section \ref{s4} presents numerical results, including the energy spectra under OBCs and PBCs as well as the topological phase diagram. Finally, we conclude in Sec.~\ref{s7}. Some derivation details are presented in Appendices.

\section {Model and spin decomposition}\label{s2}

\begin{figure}[t!]
    \centering     \includegraphics[width=1\linewidth]{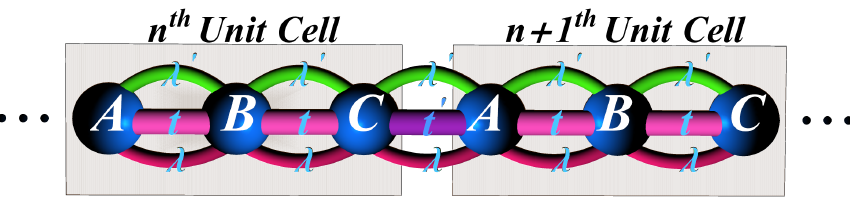}
    \caption{(Color online) Schematic representations of SSH$_3$ lattice with three sublattices per unit cell. The intracell ($t$) and intercell ($t^{\prime}$) hopping parameters are indicated by flat lines. The forward ($\lambda$) and backward ($\lambda^{\prime}$) Rashba spin-orbit couplings are represented by curved lines.}
    \label{fig1}
\end{figure}

We consider a one-dimensional multiple-sublattice SSH model comprising of three sublattices per unit cell (SSH$_3$) subjected to RSOC \cite{SSH3}, as shown in Fig.~\ref{fig1}. The system consists of a trimerized chain with Hermitian alternating intracell and intercell hopping amplitudes, and nonreciprocal nearest-neighbor RSOC terms that couple opposite spins. The tight-binding Hamiltonian of the system is given by \cite{SSH3}
\begin{eqnarray}
H &=& H_{SSH_3}+H_{RSO},
\label{e1}
\end{eqnarray}
with 
\begin{align}
H_{SSH_3} &= \sum_{\sigma}\Bigg[\sum_{n=1}^N t\,(A_{n,\sigma}^\dagger B_{n,\sigma}+B_{n,\sigma}^\dagger C_{n,\sigma}) \nonumber\\
&\qquad +\sum_{n=1}^{N-1} t'\,A_{n+1,\sigma}^\dagger C_{n,\sigma}\Bigg]+{\rm h.c.},\label{eq:H_SSH_3}
\end{align}
\begin{align}
H_{RSO} &= \sum_{\sigma }\Bigg[\sum_{n=1}^N \Big( \lambda\,(A_{n,\sigma}^\dagger B_{n,-\sigma}+B_{n,\sigma}^\dagger C_{n,-\sigma}) \nonumber \\
&+\lambda'(B_{n,\sigma}^\dagger A_{n,-\sigma}+C_{n,\sigma}^\dagger B_{n,-\sigma})\Big) \nonumber \\ 
&+\sum_{n=1}^{N-1} \Big(\lambda A_{n+1,\sigma}^\dagger C_{n,-\sigma}+\lambda' C_{n+1,\sigma}^\dagger A_{n,-\sigma})\Big)\Bigg],
\label{eq:H_SSH_RSO}
\end{align}
where $X_{n,\sigma}^\dagger$ ($X_{n,\sigma}$) is the creation (annihilation) operator on sublattice $X (\equiv A,B,C)$ of the $n$th unit cell with spin $\sigma(=\uparrow,\downarrow$), and $N$ is the number of unit cells. $t(t')=t_0+(-)\delta t$
is the intracell (intercell) hopping between nearest neighbors, and $\delta t$ is the dimerization strength. 
We assume a non-Hermitian RSOC as $\lambda(\lambda')=\lambda_0+(-)\delta\lambda$, where $\delta\lambda$ introduces nonreciprocity to the RSOC. Here, $\lambda$ and $\lambda'$ denote the forward and backward non-Hermitian Rashba
spin-orbit coupling amplitudes, respectively. 

Despite the mixture between different spin components, we note that the Hamiltonian in Eq.~\eqref{e1} preserves a spin-flip symmetry,
\begin{eqnarray}\label{eq: spin-rot sym}
[H,\mathcal{F}]=0,\qquad\mathcal{F} = \sigma_x \otimes I_{3N},
\end{eqnarray}
where $\mathcal{F}$ is the operator of spin-flip symmetry with $I_{3N}$ and $\sigma_x$ being the identity matrix with size $3N$ and the $x$-component of the Pauli matrices, respectively. Therefore, the full Hamiltonian can be decomposed into two independent sectors in the basis of the operator $\mathcal{F}$, i.e.,
\begin{eqnarray}\label{eq:Block Hamiltonian}
    H = \bigoplus_{\alpha}H_{\alpha},
\end{eqnarray}
where $\alpha=+1$ ($\alpha=-1$) are eigenvalues of $\mathcal{F}$, corresponding to the $\sigma=\uparrow$ ($\sigma=\downarrow$) sector along the $x$-direction. 
Each block $H_{\alpha}$ is given by
\begin{eqnarray}\label{eq: HamOBC}
H_{\alpha} =
\begin{pmatrix}
0 & f_{3 \alpha } &  &  &  & \\
f_{1\alpha } & 0 & f_{3\alpha } & & O  &  \\
& f_{1\alpha } & 0 & f_{4\alpha } &  & \\
& & f_{2\alpha } & 0 & \ddots  \\
& O & & \ddots &0 & \\
& & & &
\end{pmatrix}_{3N\times 3N},
\label{HamilAlpha}
 \end{eqnarray}
where $f_{1\alpha} = t +\alpha\lambda$, 
$f_{2\alpha} = t' +\alpha \lambda$, 
$f_{3\alpha}  = t +\alpha \lambda'$, and 
$f_{4\alpha} = t' +\alpha \lambda'$.  To simplify later expressions, we further define
$f_{5\alpha} = f_{1\alpha}f_{3\alpha} $ and $f_{6\alpha} = f_{2\alpha}f_{4\alpha}$.

This block-diagonal structure enables an independent analysis of spin-resolved NHSE and $\mathcal{PT}/\mathcal{APT}$ symmetry breaking within each sector, where each subsystem of $H$ behaves as a generalized non-Hermitian SSH$_3$ model with real parameters, satisfying $H_{\alpha}=H_{\alpha}^\star$. 

In Table \ref{tab:cases}, we summarize the different phases regarding the $\mathcal{PT}$ and $\mathcal{APT}$ symmetries. In particular, we identify four phases with distinct symmetry-breaking and symmetry-preserving conditions for the bulk bands, denoted as Phases I, II, IIIr, and IIIi. When the topological edge states are further taken into account, another phase with $\mathcal{PT}$/$\mathcal{APT}$-unbroken edge states emerges indicated by Phases IV and V.
The origin and transition between these phases are analyzed in detail in the following sections.

\begin{table}[b]
\centering
\caption{
Classification of regimes for each block $H_{\alpha} $ in terms of symmetry conditions.
The first four phases correspond to bulk bands, and phases IV and V correspond to edge states.}
\label{tab:cases}
\begin{tabular}{l|l|l}
\hline
 Phase&Condition & Interpretation  \\
\hline
I&$\begin{array}{l}
f_{6\alpha}<0 \land f_{5\alpha}<0,\\
\text{or } f_{5\alpha}>0 \land f_{6\alpha}< -8f_{5\alpha}
\end{array}$
& Full $\mathcal{APT}$-unbroken bulk \\
II&
$\begin{array}{l}
f_{6\alpha}>0 \land f_{5\alpha}>0,\\
\text{or } f_{5\alpha}<0 \land f_{6\alpha}> -8f_{5\alpha}
\end{array}$
& Full $\mathcal{PT}$-unbroken bulk \\
IIIr&
$f_{5\alpha}<0 \land 0<f_{6\alpha}<-8f_{5\alpha}$
& $\begin{array}{c}
\ \qquad \pm (E,E^*),\\
\ \text{real central band}
\end{array}$ \\
IIIi&
$f_{5\alpha}>0 \land -8f_{5\alpha}< f_{6\alpha}<0$
& $\begin{array}{c}
\pm (E,E^*),\\
\text{imaginary central band}
\end{array}$
 \\
\hline
IV&
$|f_{6\alpha}|>|f_{5\alpha}|,\ f_{5\alpha}<0$
& $\mathcal{APT}$-unbroken edge\\
V&
$|f_{6\alpha}|>|f_{5\alpha}|,\ f_{5\alpha}>0$
& $\mathcal{PT}$-unbroken edge \\
\hline
\end{tabular}
\end{table}

\section{Symmetry and Spectral Phase Classification}
\label{s3}

An important characteristic in non-Hermitian systems is the contrast between the spectra of PBC and OBC. While a PBC system consists of bulk bands only described by the Bloch Hamiltonian, an OBC system breaks translational invariance, thus allowing topological edge states and NHSE to emerge, where the latter greatly influences the bulk band structure. 
Therefore, to unveil the spectral features of both bulk and topological edge states, here we focus on symmetries acting on the OBC Hamiltonian.

\subsection{$\mathcal{PT}$ and $\mathcal{APT}$ symmetries and bulk phases}\label{BulkSolve}

We define the $\mathcal{PT}$ symmetry via the operator 
\[
\mathcal{PT}=U_{PT}\mathcal{K},
\] where the unitary part $U_{PT}=I$ is the identity matrix and $\mathcal{K}$ is the complex conjugate operator. This symmetry ensures that
\[\mathcal{PT}H_\alpha(\mathcal{PT})^{-1}=H_\alpha.\]

Similarly, the system also exhibits an $\mathcal{APT}$ symmetry, satisfying
\[\mathcal{APT}H_\alpha(\mathcal{APT})^{-1}=-H_\alpha.\]
The corresponding operator is
\[\mathcal{APT}=U_{APT}\mathcal{K},\]
where
\begin{eqnarray}\label{e3}
U_{APT} &=&
\begin{pmatrix}
 1&   &        &   &  \\
 & -1 &        & O &   \\
 &   & \ddots &   &   \\
 & O &        &  &   \\
  &   &      &   &   
\end{pmatrix}_{3N\times 3N}.
\end{eqnarray}
Note that the two symmetries defined here do not involve the spatial inversion operator. We refer to them as the $\mathcal{PT}$ and $\mathcal{APT}$ symmetries because they play analogous roles in constraining the spectral features of the Hamiltonian. Consequently, the eigenenergies are restricted to the following phases:
\begin{itemize}
    \item Phase I ($\mathcal{APT}$-unbroken): Bulk eigenvalues are purely imaginary;
    \item Phase II ($\mathcal{PT}$-unbroken): Bulk eigenvalues are purely real;
    \item Phase III (both symmetries broken): Bulk eigenvalues form quartets $(E,-E,E^*,-E^*)$.
\end{itemize}
Furthermore, we note that symmetry-broken and symmetry-unbroken conditions may coexist within different energy regimes. In particular, in our three-band model, we find that the symmetry-broken quartets are formed from the lowest- and highest-energy bands (ordered by their real energies). In comparison, the second (central) band has either purely real ($\mathcal{PT}$-unbroken) or purely imaginary ($\mathcal{APT}$-unbroken) energies. In Table \ref{tab:cases}, we denote these mixed configurations as IIIr or IIIi phases, respectively (see Appendix \ref{App:PBCsolution} for details). In the following, we refer to the first and second bands as the lowest and central bands, indexed by "1" and "2", respectively. 

To determine the OBC bulk phase boundaries, we employ the non-Bloch band theory~\cite{NonBloch1,NHSE1,NHSE2,NonBloch4,NonBloch5}. 
Taking Fourier transform from (\ref{e1}), the generalized non-Bloch Hamiltonian of subsystem $\alpha$ can be obtained 
as (see Appendix~\ref{app:beta}),
\begin{equation}
h_{\alpha }(\beta_{\alpha})=
\begin{pmatrix}
0 &f_{3\alpha } & f_{4\alpha } \beta_{\alpha}^{-1}\\
f_{1\alpha }&0 & f_{3\alpha }\\
f_{2\alpha } \beta_{\alpha}&f_{1\alpha }&0
\end{pmatrix},
\label{Hbeta}
\end{equation}
where the GBZ parameter is
\begin{equation}
\beta_{\alpha }=
\frac{|f_{1\alpha }|\sqrt{f_{4\alpha }}}
{|f_{3\alpha }|\sqrt{f_{2\alpha }}}
e^{ik}.
\label{eq: beta}
\end{equation}

Substituting Eq.~(\ref{eq: beta}) into Eq.~(\ref{Hbeta}) yields the cubic secular equation
\begin{equation}
-E^3+p_\alpha E+q_\alpha=0,
\label{Eq:CubicEquation}
\end{equation}
with
\begin{equation}
\begin{aligned}
p_{\alpha } &= 2f_{5\alpha }+f_{6\alpha }, \\
q_{\alpha } &= 2 i^{\frac{2-\operatorname{Sign}(f_{2\alpha})-\operatorname{Sign}(f_{4\alpha})}{2}}
|f_{5\alpha }|\sqrt{|f_{6\alpha }|}\cos k.
\end{aligned}
\end{equation}

The bulk spectral phases follow directly from the properties of the cubic equation. For $f_{6\alpha}<0$, $q_\alpha$ is purely imaginary, giving either a purely imaginary spectrum (Phase I), when
\begin{equation}
f_{5\alpha}<0,\quad \text{or}\quad
(f_{5\alpha}>0,\;f_{6\alpha}<-8f_{5\alpha}),
\label{eq:phase1}
\end{equation}
or a mixed spectrum with one imaginary eigenvalue and one complex-conjugate pair (Phase IIIi), when
\begin{equation}
f_{5\alpha}>0,\qquad
-8f_{5\alpha}<f_{6\alpha}<0.
\label{eq:phase3i}
\end{equation}
For $f_{6\alpha}>0$, $q_\alpha$ is real. The spectrum is purely real (Phase II), when
\begin{equation}
f_{5\alpha}>0,\quad \text{or}\quad
(f_{5\alpha}<0,\;f_{6\alpha}>-8f_{5\alpha}).
\label{eq:phase2}
\end{equation}
Whereas it enters Phase IIIr, consisting of one real eigenvalue and one complex-conjugate pair, for
\begin{equation}
f_{5\alpha}<0,\qquad
0<f_{6\alpha}<-8f_{5\alpha}.
\label{eq:phase3r}
\end{equation}
The derivation of these phase boundaries are given in Appendix~\ref{App:PBCsolution}.
The analytical expressions for the phase boundaries, obtained from the non-Bloch Hamiltonian,  completely determine the OBC bulk phase diagram.
 
In the following section, they will be compared with the numerical spectra.

\subsection{Analytical characterization of edge spectra}\label{EdgeSolve}

To characterize the edge spectrum under OBCs, we introduce a similarity transformation that maps the non-Hermitian Hamiltonian to an equivalent Hermitian or anti-Hermitian Hamiltonian while preserving its eigenvalues. The transformation matrix is defined as
\begin{eqnarray}
[S_{\alpha}]_{n+1,n+1}
=
r_{1\alpha}^{\frac12\left(n-\lfloor n/3\rfloor\right)}
r_{2\alpha}^{\frac12\lfloor n/3\rfloor},
\label{e4}
\end{eqnarray}
where $r_{1\alpha}=f_{1\alpha}/f_{3\alpha}$ and
$r_{2\alpha}=f_{2\alpha}/f_{4\alpha}$.
The transformed Hamiltonian can be obtained by
\begin{equation}
H'_{\alpha}=S_\alpha^{-1}H_\alpha S_\alpha,
\label{Eq:SimHamiltonian}
\end{equation}
which maps the original non-Hermitian Hamiltonian to a spectrally equivalent form in which the conventional bulk--boundary correspondence is recovered. Depending on the signs of $f_{5\alpha}$ and $f_{6\alpha}$, the transformed Hamiltonian becomes either Hermitian or anti-Hermitian (See Appendix~\ref{App1} for more details),
\begin{equation}
H'_{\alpha}\in
\begin{cases}
AZ^\dagger~(\mathrm{AIII}), &
f_{5\alpha},f_{6\alpha}<0,\\
AZ~(\mathrm{BDI}), &
f_{5\alpha},f_{6\alpha}>0,
\end{cases}
\label{eq:AZandAZdagger}
\end{equation}
corresponding to purely imaginary and purely real spectra, respectively. Although the discrete symmetry operators remain fixed, $H'_\alpha$ itself changes character across this transition: when $f_{5\alpha}, f_{6\alpha}>0$, $S_\alpha$ has real diagonal entries, making $H'_\alpha$ Hermitian with a real spectrum; when $f_{5\alpha}, f_{6\alpha}<0$, these entries become alternately real and imaginary, making $H'_\alpha$ anti-Hermitian with an imaginary spectrum. The transition between AZ and AZ$^\dagger$ is thus a property of the transformation $S_\alpha$, not of the underlying symmetry.

Assuming exponentially localized edge states,
\[
(\psi_A,\psi_B,\psi_C)^T\xi_\alpha^n,
\]
the boundary Schrödinger equation yields (see Appendix~\ref{App1})
\begin{align}
E_{edge,\alpha}&=\pm T_{1\alpha},\quad
\xi_\alpha=-\frac{T_{3\alpha}}{T_{1\alpha}}.
\label{eq:E_edge_lambda}
\end{align}
where $T_{1\alpha} = \operatorname{Sign}(f_{3\alpha})i^{\frac{1-\operatorname{Sign}(f_{5\alpha})}{2}}
  \sqrt{|f_{5\alpha}|}$ and $T_{3\alpha} = i^{\frac{2-\operatorname{Sign}(f_{2\alpha})
  -\operatorname{Sign}(f_{4\alpha})}{2}}
  \sqrt{|f_{6\alpha}|}$. Localized edge states therefore exist when
\begin{equation}
|f_{6\alpha}|>|f_{5\alpha}|.
\label{ProCond}
\end{equation}
Since $T_{1\alpha}$ is real for $f_{5\alpha}>0$ and purely imaginary for $f_{5\alpha}<0$, the edge spectrum is classified into two phases:
\begin{align}
\text{Phase IV:}\quad
&|f_{6\alpha}|>|f_{5\alpha}|,\qquad
f_{5\alpha}<0,
\label{Eq:APT}
\\
\text{Phase V:}\quad
&|f_{6\alpha}|>|f_{5\alpha}|,\qquad
f_{5\alpha}>0.
\label{Eq:PT}
\end{align}

Note that although the $\mathcal{PT}/\mathcal{APT}$ symmetries determine whether the edge modes possess real or imaginary energies, their existence and robustness are ultimately protected by the underlying Bernard–LeClair symmetries of the Hamiltonian. We investigated the symmetries of edge states and  their robustness against various types of perturbations in Appendix~\ref{app: EdgeSymmetry}.

Together with the bulk phase boundaries derived above, these analytical results completely determine the symmetry-resolved bulk and edge spectra. The resulting phases are summarized in Table \ref{tab:cases} and the corresponding phase diagram will be discussed in the following subsection. 

\begin{figure}[t!]
    \centering     \includegraphics[width=1\linewidth]    {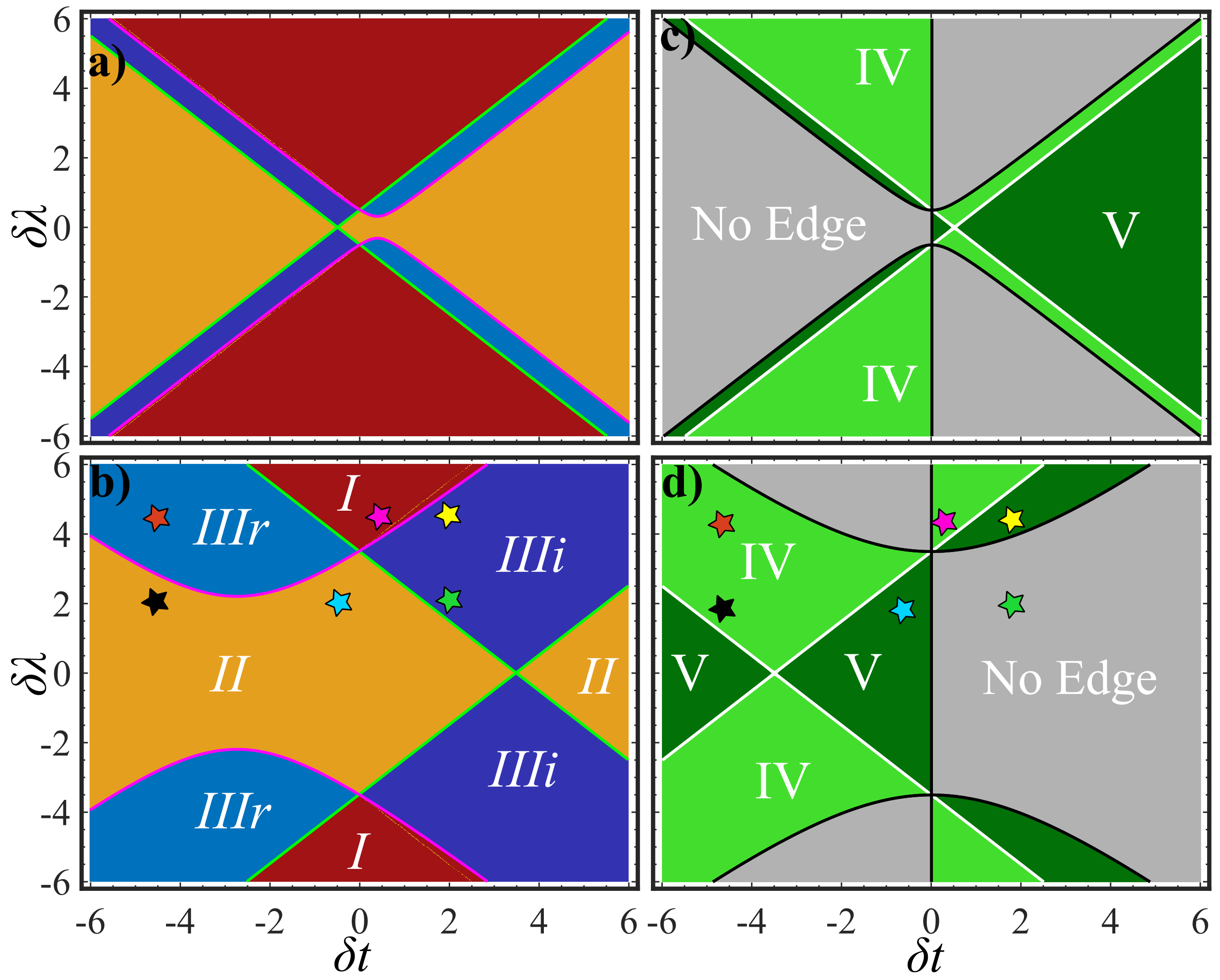}
    \caption{(Color online) Phase diagram of the spin-down (upper row) and spin-up (lower row) sectors in the ($\delta t,\delta\lambda$) plane for the bulk (left column) and edge (right column) states. The green line separates the regions with $q_{\alpha}\gtrless 0$, while the magenta line separates the mixed states from others. Also, the boundary between phases IV and V and the localized and delocalized edge states are indicated in white and black, respectively. Star markers denote points at $\delta t=-4.5,-0.5,0.5,1.9$ and $\delta\lambda=2,4.5$.  Here, $t_0=1.5$, $\lambda_0=2$.   
    }
    \label{fig:Diagram_phase}
\end{figure}

\begin{figure}[t!]
    \centering     \includegraphics[width=1\linewidth]    {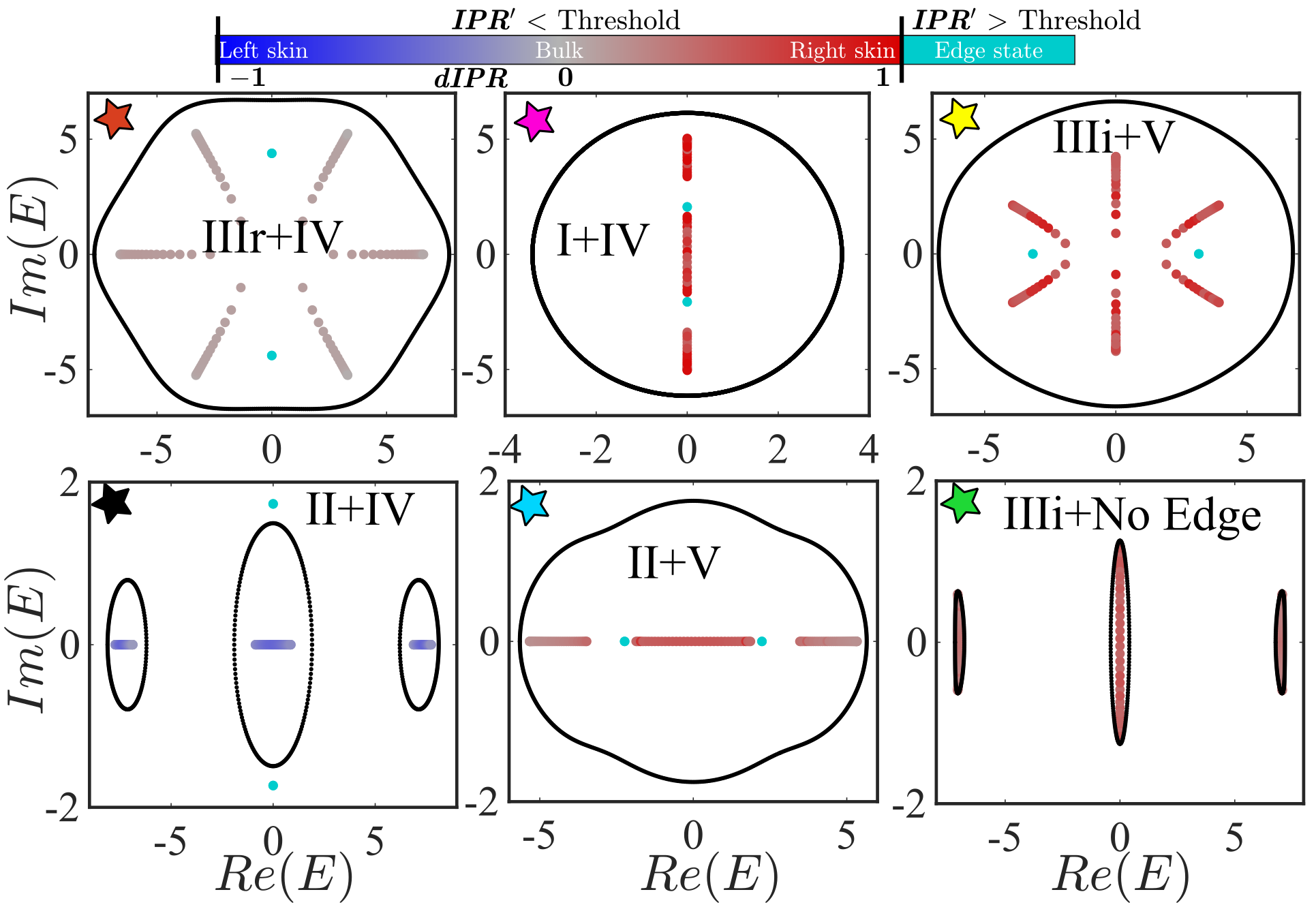}
    \caption{(Color online) Representative spectra of the spin-up ($H_+$) sector in the complex energy plane, for parameter values marked by the colored stars in Figs. \ref{fig:Diagram_phase}(b) and \ref{fig:Diagram_phase}(d), each corresponding to a specific phase listed in Table \ref{tab:cases}. The PBC spectrum is shown as a solid block loop enclosing the spectral region of the bulk Hamiltonian. The OBC spectrum is shown as colored points. Eigenstate localization is encoded by IPR$'$ and dIPR: localized edge states (cyan, IPR$'$ above threshold 0.1),  left-localized skin modes (blue, dIPR $\rightarrow -1$), right-localized skin modes (red, dIPR $\rightarrow 1$), and extended bulk modes (grey, dIPR $\approx 0$).
    }
    \label{fig:ComplexPlane}
\end{figure}

\subsection{Phase diagram of bulk and edge spectra}

Figure~\ref{fig:Diagram_phase} summarizes the analytically obtained phase diagrams of the bulk [Figs.~\ref{fig:Diagram_phase}(a,b)] and edge [Figs.~\ref{fig:Diagram_phase}(c,d)] spectra. Since the two spin sectors are decoupled by the $\mathcal{F}$ symmetry [Eq.~(\ref{eq: spin-rot sym})], they evolve independently in the $(\delta t,\delta\lambda)$ parameter space, giving rise to spin-resolved symmetry transitions. The top (bottom) row is for the spin-down (spin-up) sector.

The bulk phase diagrams (left panels) contain four spectral phases. Phase II (yellow regions) corresponds to the $\mathcal{PT}$-unbroken regime with purely real spectra, whereas Phase I (red regions) represents the $\mathcal{APT}$-unbroken regime with purely imaginary spectra. The remaining regions (IIIr and IIIi, shown in blue) are mixed phases in which one band remains on the symmetry axis while the other two form a complex-conjugate pair. The phase boundaries follow directly from the analytical conditions derived in Secs. \ref{BulkSolve} and \ref{EdgeSolve}. The green line ($f_{6\alpha}=0$) separates regions with real and imaginary values of $q_\alpha$. While the magenta line ($f_{6\alpha}=-8f_{5\alpha}$) marks the transition between purely real/imaginary spectra and the mixed phases. 

The edge phase diagrams (right panels) consist of two localized phases: Phase IV (light green regions), where the edge states possess purely imaginary energies, and Phase V (dark green regions), where the edge-state energies are purely real. The gray regions indicate parameter regimes in which the localization condition $|f_{6\alpha}|>|f_{5\alpha}|$ is violated and no localized edge states exist. For the edge phase boundaries, the white line separates Phases IV and V, whereas the black line denotes the localization transition.

One of the most interesting features of Fig.~\ref{fig:Diagram_phase} is the coexistence of different symmetry sectors in the bulk and edge spectra. In particular, the parameter regime satisfying
$f_{5\alpha}<0$,
$f_{6\alpha}>-8f_{5\alpha}$,
and
$|f_{6\alpha}|>|f_{5\alpha}|$
supports a fully $\mathcal{PT}$-unbroken bulk (Phase II) together with $\mathcal{APT}$-unbroken edge states (Phase IV). This bulk--edge symmetry separation contrasts with the conventional scenario of Ref.~\cite{wu2025topological}, where $\mathcal{APT}$ symmetry produces imaginary bulk bands beneath real edge states. Conversely, when $f_{5\alpha},f_{6\alpha}>0$ or $f_{5\alpha},f_{6\alpha}<0$, both the bulk and edge spectra belong to the AZ (BDI) or AZ$^\dagger$ (AIII) symmetry class, respectively, realizing globally $\mathcal{PT}$- or $\mathcal{APT}$-symmetric phases. These results establish the complete spin-resolved non-Hermitian bulk--boundary correspondence of the SSH$_3$-Rashba model.

\begin{figure}[t]
    \centering     \includegraphics[width=1\linewidth]{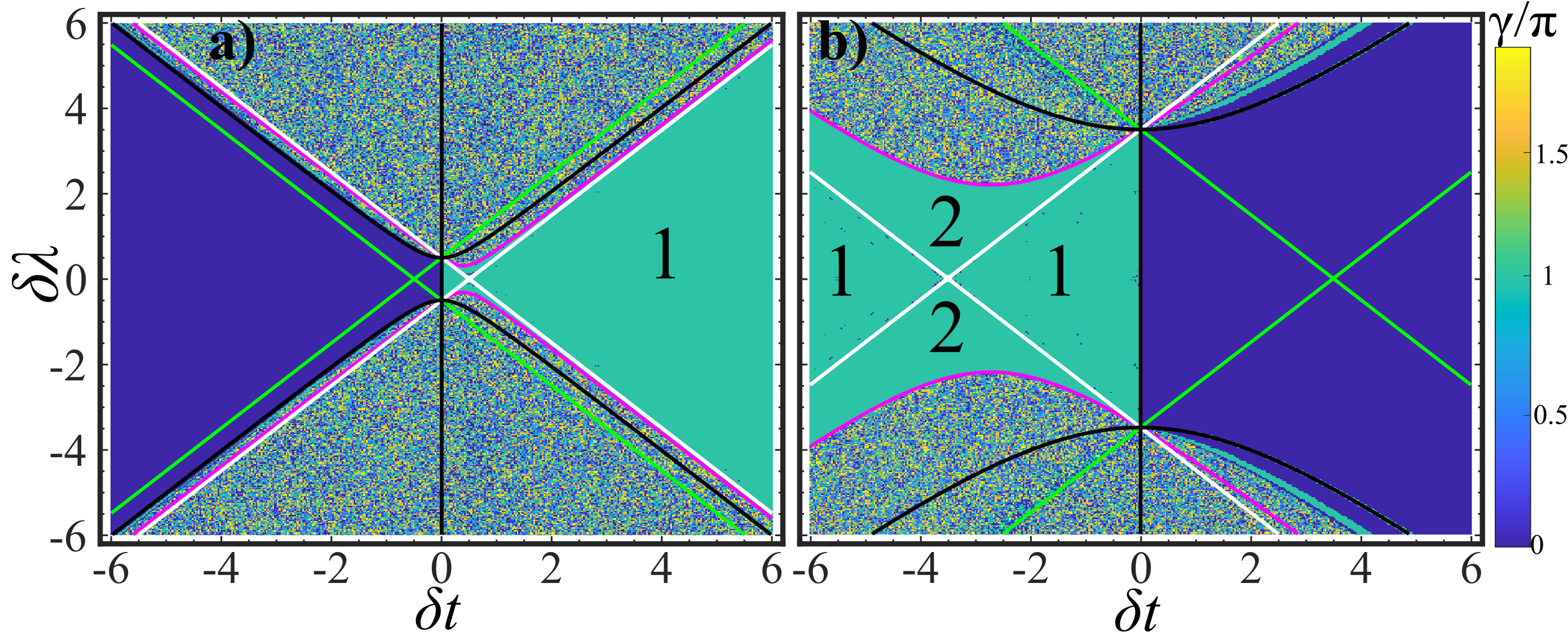}
    \caption{(Color online)  Berry phase diagram in the $(\delta t, \delta\lambda)$ plane for (a) the spin-down and (b) the spin-up sectors exhibiting a strong asymmetry. The regions labeled $"1"$ and $"2"$ represent sub-regions where the quantized Berry phase is carried by the lowest and central energy bands, respectively. Here, $t_0=1.5$ and $\lambda_0=2$. 
    }
    \label{fig:BerryDiagram_phase}
\end{figure}

\section{Numerical Verification of the Analytical Results} \label{s4}

\subsection{Analytical tools for phase characterization}

To systematically characterize the interplay of non-Hermiticity, topology, and localization, we employ two approaches: localization metrics for finite systems under OBC and a topological invariant derived from the momentum-space Hamiltonian.

\textbf{Localization Metrics:} In non-Hermitian systems with open boundaries, the NHSE can obscure topological edge states. As mentioned in Sec. \ref{s2}, we invoke a similarity transformation to map the Hamiltonian $H_\alpha$ and its eigenstate $|\psi_\alpha\rangle$ to $H'_\alpha = S_\alpha^{-1} H_\alpha S_\alpha$ and $|\psi'_\alpha\rangle = S_\alpha^{-1}|\psi_\alpha\rangle$, respectively, where $|\psi_\alpha'\rangle$ is an eigenstate of $H'_\alpha$ with the same eigenvalue. It is well known that, in the transformed basis, the NHSE is removed, leading to boundary-localized states of $H_{\alpha}$ appearing as extended ones. We calculate the inverse participation ratio ($\text{IPR}'$)~\cite{IPR} 
\begin{equation}
\text{IPR}'_n = \frac{\sum_i |\psi'_{n,i}|^4}{\left(\sum_i |\psi'_{n,i}|^2\right)^2},
\end{equation}
(where the prime denotes the IPR of $H'$), using the eigenstates $\psi'_{n,i}$ of $H'$ [Hamiltonian~(\ref{Eq:SimHamiltonian})]. States with an ${\rm IPR}'$ larger than a specific threshold (colored cyan in our subsequent spectral plots) are unambiguously identified as topological edge modes. The remaining bulk and skin states are colored according to their directional IPR (dIPR)~\cite{dirIPR1,dirIPR2,dirIPR3} which lies in $[-1,1]$, given by
\begin{equation}
\text{dIPR}_n = \text{sgn}(x_n) \frac{\sum_i |\psi_{n,i}^R|^4}{\left(\sum_i |\psi_{n,i}^R|^2\right)^2},
\end{equation}
where $\psi^R_{n,i}$ is the right eigenstate of the $H$ [Hamiltonian~(\ref{eq: HamOBC})] and the sign is determined by the center-of-mass offset
\begin{equation}
x_n = \sum_i \left(i - \frac{N}{2} - \delta\right) |\psi_{n,i}^R|^2,
\end{equation}
with $\delta = 0.1$ being a small offset to avoid ambiguity at the exact center. The signed dIPR$_n$ allows us to distinguish left-localized skin states (dIPR$_n < 0$), right-localized skin states (dIPR$_n > 0$), and extended or purely localized bulk states ($|\text{dIPR}_n| \sim 0$).

\textbf{Topological invariant:} 
The topological properties of the model are characterized by the geometric phase acquired by its eigenstates. In non-Hermitian systems, the topological invariant must be defined in terms of biorthogonal eigenstates. The biorthogonal Berry phase associated with band $n$ in one dimension is given by~\cite{berry1,berry2}
\begin{align}
\gamma_n = i \int_{\mathrm{GBZ}} \langle u^L_n(k) \,|\, \partial_k u^R_n(k) \rangle \, dk,    
\end{align}
where $|u^{R}_n(k)\rangle$ and $|u^{L}_n(k)\rangle$ are the right and left eigenvectors of Hamiltonian~(\ref{Hbeta}) and the integration is performed over the GBZ for characterizing topological edge states under OBCs.

\subsection{Spectral signatures and topological invariants}

Figure~\ref{fig:ComplexPlane} displays the complex-energy spectra under both OBCs and PBCs for representative parameter points indicated by the colored stares in Fig. \ref{fig:Diagram_phase}. While the PBC spectrum forms closed loops (solid black lines), the OBC spectra (colored dots) collapses onto line segments (residing on the real, imaginary, or oblique axes) depending on the phase. This deviation between PBC and OBC spectra confirms the ubiquitous NHSE and the breakdown of conventional bulk-boundary correspondence across the phase diagram.

The black and cyan stars in Fig. \ref{fig:ComplexPlane} mark two examples of $\mathcal{PT}$-unbroken bulk phases, with distinct edge spectral properties. The black-star panel, in particular, reveals our central result: the bulk states (blue dots, representing left-localized skin modes) remain purely real (Phase II), whereas the edge states (cyan dots) are pinned to the imaginary axis (Phase IV). This coexistence of a fully real bulk with purely imaginary edge states constitutes an \textit{edge-only unbroken $\mathcal{APT}$ symmetry breaking}. The cyan-star panel demonstrates a configuration where both the Phase V edge states and the Phase II bulk states are coexist and share purely real energies.

An example of $\mathcal{APT}$-unbroken bulk phase is shown by the magenta star of Fig.~\ref{fig:ComplexPlane}, where
bulk states manifest as right-localized skin modes confined to the imaginary axis, confirming that the NHSE persists in this regime. Furthermore, the edge-state energies reside symmetrically within the imaginary gaps.

Finally, the red, yellow, and green stars of Fig.~\ref{fig:ComplexPlane} demonstrate three typical examples of phase III where both $\mathcal{PT}$ and $\mathcal{APT}$ symmetries are broken for the bulk, yielding complex eigenvalues. The bulk states form right-localized skin modes that exhibit distinct structures in the complex-energy plane. The energetic behaviors of the central band, whether being real (Phase IIIr, red star) or imaginary (Phase IIIi, yellow and green stars), provide a direct spectroscopic signature to distinguish these phases, demonstrating that bulk and edge states can belong to fundamentally different symmetry classes.

Figure~\ref{fig:BerryDiagram_phase} shows the biorthogonl Berry phase in the $(\delta t,\delta\lambda)$ parameter space for the spin-down [Fig.~\ref{fig:BerryDiagram_phase}(a)] and spin-up [Fig.~\ref{fig:BerryDiagram_phase}(b)] sectors, illustrating strongly spin-dependent topological features. The diagrams exhibit three distinct regimes: (i) topological regions characterized by a quantized Berry phase of $\pi$ (green areas), which correspond to Phases IV and V, where edge states emerge with gapped bulk spectra;
(ii) trivial regions with a zero Berry phase (dark blue areas), where no edge modes exist despite the spectrum being gapped; and (iii) speckled regions where the Berry phase becomes ill-defined due to band-touching points that close the spectral gap and prevent band separation. The phase boundaries plotted over the phase diagram perfectly match with those derived for the bulk [Eqs. (\ref{eq:phase1}-\ref{eq:phase3r})] and edge states [Eqs. (\ref{Eq:APT}-\ref{Eq:PT})].

Remarkably, these phase diagrams show a pronounced asymmetry between the two spin sectors with respect to the trimerization parameter $\delta t$. This asymmetry originates from the spin-orbit coupling, which acts as an effective spin-dependent potential. It explicitly breaks the symmetry between the spin components, shifting the gap-closing conditions for the topological phase transition to opposite signs of $\delta t$ for each spin species. Consequently, in the spin-down sector [Fig. \ref{fig:BerryDiagram_phase}(a)], the topological $\pi$-phase is confined to the $\delta t>0$ half-plane, where the non-trivial Berry phase is carried by the lowest band ("1").

In contrast, the spin-up sector [Fig. \ref{fig:BerryDiagram_phase}(b)] reveals its topological phase when $\delta t<0$. Furthermore, the structure of the spin-up topological region is notably more complicated. It is partitioned by internal crossing boundaries into distinct sub-regions, where the quantized Berry phase is carried by either the lowest band ("1") or the central band ("2"). This structure results from the interplay between the three-sublattice feature of the underlying lattice and non-Hermitian effects. These internal boundaries set specific lines where additional gap closings and non-Hermitian band crossings take place within the gapped topological phase. Subsequently, as parameters cross these internal boundaries, a band inversion occurs, transferring topological charge between the lowest and the central bands.

\begin{figure*}[t!]
    \centering     \includegraphics[width=1\linewidth]{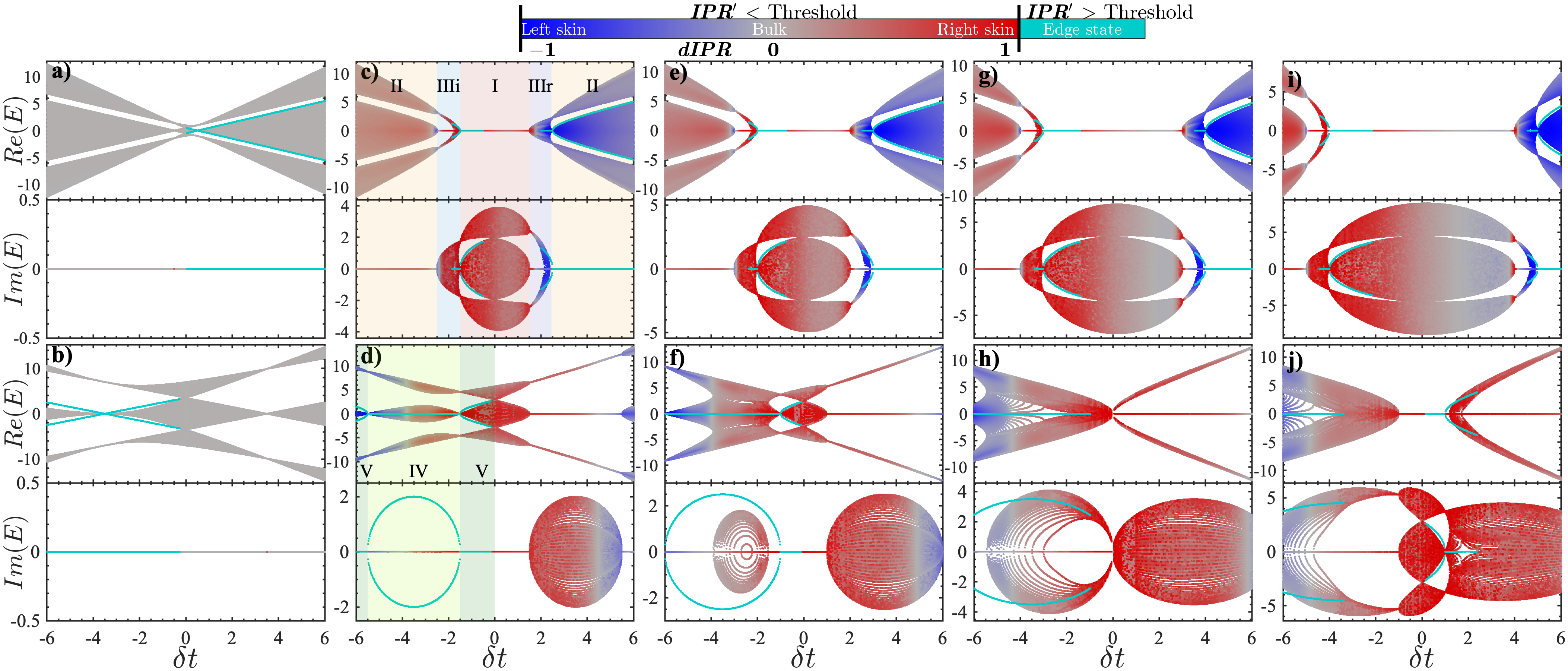}
    \caption{(Color online) Energy levels as a function of $\delta t$, where the color coding represents the IPR$'$ and dIPR. The top (bottom) row corresponds to the spin-down (spin-up) subspace. Each column corresponds to a different value of $\delta \lambda$: $0$, $2$, $2.5$, $3.5$, and $4.5$ from left to right. In panels (c) and (d), the bulk phase labels (I, II, IIIr, IIIi) and edge phase labels (IV, V) are marked, respectively, only to avoid visual clutter, even though both classifications apply throughout. The IPR$'$ threshold for determining topological edge states (cyan) is set to 0.035 for all panels. Here, $t_0=1.5$ and $\lambda_0=2$.
    }
\label{fig:energy_levels_IPR_DIPR}
\end{figure*}
\subsection{Open-boundary spectra and symmetry transitions}

To provide a more detailed analysis, Figure \ref{fig:energy_levels_IPR_DIPR} displays the real and imaginary parts of the energy spectra for the spin-down ($H_{-}$) and spin-up ($H_{+}$) subsystems (top and bottom rows, respectively) as a function of the trimerization parameter $\delta t$ under OBCs. The color mapping encodes the IPR$'$ and dIPR. From the left to the right column, the non-reciprocal RSOC parameter $\delta\lambda$ increases, taking values of 0, 2, 2.5, 3.5, and 4.5, respectively.

In the Hermitian limit at $\delta\lambda=0$ [Figs. \ref{fig:energy_levels_IPR_DIPR}(a) and \ref{fig:energy_levels_IPR_DIPR}(b)], the imaginary part of the spectrum vanishes. The bulk bands exhibit a typical $SSH_3$-like dispersion \cite{SSH3}. Each subsystem hosts maximally localized edge states (cyan points) that emerge within the band gaps, appearing specifically at $\delta t>0$ for $H_{-}$ [Fig. \ref{fig:energy_levels_IPR_DIPR}(a)] and at $\delta t<0$ for $H_{+}$ [Fig. \ref{fig:energy_levels_IPR_DIPR}(b)]. These in-gap edge states possess a finite-energy dispersion before merging into the bulk continuum at larger $|\delta t|$.

As $\delta\lambda$ increases [Figs. \ref{fig:energy_levels_IPR_DIPR}(c)–(j)], the non-reciprocal spin–orbit coupling drives the system into a non-Hermitian regime. It yields complex energy eigenvalues, leading to imaginary components in the spectrum. The imaginary bands form loops as $\delta t$ varies, signaling the appearance of exceptional points and the onset of an $\mathcal{APT}$-symmetric character. The emergence of these imaginary-energy loops coincides with the formation of zero-energy flat bands in the $\text{Re(E)}$ spectrum bounded by these exceptional points. In addition, since the non-Hermiticity originates from a spin-dependent hopping asymmetry, the spectra of $H_{+}$ and $H_{-}$ show rather different behaviors as $\delta\lambda$ increases. Specifically, for $H_{-}$, the $\mathcal{PT}$-broken regime emerges near $\delta t=0$ and expands for stronger non-Hermiticity; in contrast, it initially manifests as two disconnected "domes" for $H_{+}$ which broaden toward $\delta t=0$ and eventually merge together as $\delta\lambda$ increases.

The topological edge states with IPR$'$ larger than the chosen threshold also show distinguish behaviors. In the $H_{-}$ subsystem, localized edge states now also emerge at $\delta t<0$, where they disappear in the Hermitian limit
[Fig. \ref{fig:energy_levels_IPR_DIPR}(a)]. In the subsystem $H_{+}$, while topological edge states remain restricted to $\delta t<0$, they undergo a $\mathcal{PT}$-symmetry transition, identified by their complex-conjugate imaginary energies with zero real parts [e.g., region IV in Figs. \ref{fig:energy_levels_IPR_DIPR}(d)]. 

Further increasing $\delta \lambda$, the edge states in both the real and imaginary spectra hybridize with the bulk states in certain ranges of $\delta t$, exhibiting delocalized feature [e.g., Figs. \ref{fig:energy_levels_IPR_DIPR}(h) and \ref{fig:energy_levels_IPR_DIPR}(j)]. Physically, this hybridization is a consequence of the breakdown of mode orthogonality induced by asymmetric RSOC-driven hopping, accumulating all eigenmodes toward one edge. The resulting coalescence of eigenvalues and eigenvectors near the exceptional points creates a continuous spectral connection between the topological edge states and bulk continuum.

Interestingly, the imaginary spectra of Fig. \ref{fig:energy_levels_IPR_DIPR}(j), reveal that across $\delta t=0$, the imaginary gap closes and reopens as a direct result of the merging of exceptional-domes. A topological phase transition is seen at this point, signaled
by the appearance of topological edge states at finite $\delta t >0$. Note that the bulk spectrum in this region remains in the $\mathcal{APT}$-symmetric regime on both sides of $\delta t=0$. 

\section {Summary} \label{s7}

We investigated the topological properties, spectral phases, and localization behaviors of a one-dimensional non-Hermitian trimerized SSH lattice. The non-Hermiticity originates from the considered nonreciprocal RSOC. By exploiting the spin-flip symmetry, we decoupled the system into two independent spin-resolved subsystems. This decomposition revealed a pronounced spin-sector asymmetry, in which the effective non-Hermitian parameters impose distinct $\mathcal{PT}$ and $\mathcal{APT}$ symmetry-breaking phase diagrams. Subsequently, we realized spin-dependent topological boundaries and NHSE.

Furthermore, we identify a generalized bulk-boundary correspondence of symmetry-protected topological edge states for the spin-resolved subsystems [see Appendix \ref{app: EdgeSymmetry} for more details], together with a uniquely spin-dependent NHSE for bulk states. 
We also 
characterized 
the features of the localized topological edge states 
via analytical derivations of the GBZ and numerical calculations of the energy spectra. These approaches uncovered a regime of bulk-edge symmetry class separation; a fully $\mathcal{PT}$-symmetric (AZ/BDI) bulk coexists with fully $\mathcal{APT}$-symmetric (AZ$^\dagger$/AIII) edge states. 

Our model thus provides a simple, yet remarkably rich, platform for exploring spin-resolved non-Hermitian topology, subsystem skin effects, and parameter-driven spectral transitions. These findings pave the way for future engineering of spin-resolved topological states in open quantum systems.

\section*{Data Availability Statement}
The data that support the findings of this study are available from the authors upon reasonable request.

\section*{Acknowledgement}
We are grateful to D. Bauer and L. E. Foa Torres for useful comments. M.J. gratefully
acknowledges funding by the Deutsche Forschungsgemeinschaft (DFG, German Research Foundation) through IRTG
2676/1 ‘Imaging of Quantum Systems’, project no. 437567992. M.V.H. gratefully acknowledges the support from the research council of the University of Zanjan.
L.Z. acknowledges support by the National Natural Science Foundation of China (Grant No.~12275260), the Shandong Provincial Natural Science Foundation (Grant No.~ZR2026QB21), and the Fundamental Research Funds for the Central Universities (Grant No. 202364008).
L.L. acknowledges support by the National Natural Science Foundation of China (Grants No. 12474159), and the Guangdong Provincial Quantum Science Strategic Initiative (Grants No. GDZX2504003 and GDZX2504006).

\clearpage

\appendix

\section{Derivation of the generalized Bloch factor $\beta_{\alpha}$}
\label{app:beta}

In this appendix, we derive the generalized Bloch factor $\beta_{\alpha}$ for each sector within the framework of non-Bloch band theory. From Eq.~(\ref{eq: HamOBC}), the Bloch Hamiltonian can be can be obtained as
\begin{equation}
H(k)=h_{+}(k)\oplus h_{-}(k),
\end{equation}
where
\begin{equation}
h_{\alpha}(k)=
\begin{pmatrix}
0 & f_{1\alpha} & f_{4\alpha}e^{-ik}\\
f_{2\alpha} & 0 & f_{3\alpha}\\
f_{3\alpha}e^{ik} & f_{4\alpha} & 0
\end{pmatrix},
\end{equation}
with the coefficients $f_{i\alpha}$ defined in the main text. To describe the energy spectrum under OBCs, the conventional Bloch phase factor is replaced by the generalized Bloch factor according to
\begin{equation}
e^{ik}\rightarrow\beta_{\alpha},\qquad
e^{-ik}\rightarrow\beta_{\alpha}^{-1},
\end{equation}
which transforms the Bloch Hamiltonian into the non-Bloch Hamiltonian
\begin{equation}
h_{\alpha}(\beta_{\alpha})=
\begin{pmatrix}
0 & f_{1\alpha} & f_{4\alpha}\beta_{\alpha}^{-1}\\
f_{2\alpha} & 0 & f_{3\alpha}\\
f_{3\alpha}\beta_{\alpha} & f_{4\alpha} & 0
\end{pmatrix}.
\end{equation}

The allowed values of the generalized Bloch factor $\beta_{\alpha}$ are determined by the characteristic equation
\begin{equation}
\det\left[h_{\alpha}(\beta_{\alpha})-EI\right]=0.
\end{equation}
Evaluating this determinant explicitly yields
\begin{equation}
f_{3\alpha}^{2}f_{2\alpha}\beta_{\alpha}^{2}
+\left(-E^{3}
+2Ef_{1\alpha}f_{3\alpha}
+Ef_{2\alpha}f_{4\alpha}\right)\beta_{\alpha}
+f_{4\alpha}f_{1\alpha}^{2}=0.
\label{CharacEq}
\end{equation}

Although $h_{\alpha}$ is a $3\times3$ Hamiltonian yielding three energy bands, Eq.~(\ref{CharacEq}) is quadratic in $\beta_{\alpha}$. This does not imply that an energy band is missing. Rather, the characteristic equation defines a constraint relating the energy $E$ and the generalized Bloch factor $\beta_{\alpha}$. Consequently, the roots $\beta_{\alpha,1}(E)$ and $\beta_{\alpha,2}(E)$ depend parametrically on the energy $E$, and the complete band structure is recovered by solving for $E$ as a function of $\beta_{\alpha}$.

Following the non-Bloch band theory for one-dimensional non-Hermitian systems under OBC, the GBZ is determined by the continuum-limit condition,
\begin{equation}
|\beta_{\alpha,1}(E)|=|\beta_{\alpha,2}(E)|.
\end{equation}
Applying Vieta's theorem to Eq.~(\ref{CharacEq}) gives
\begin{equation}
|\beta_{\alpha,1}\beta_{\alpha,2}|
=\left|\frac{c}{a}\right|
=r_{\alpha}^{2},
\end{equation}
where the coefficients are defined as
\begin{equation}
a=f_{3\alpha}^{2}f_{2\alpha},
\qquad
c=f_{4\alpha}f_{1\alpha}^{2}.
\end{equation}
Therefore, the GBZ radius is determined to be
\begin{equation}
r_{\alpha}
=\frac{|f_{1\alpha}|\sqrt{f_{4\alpha}}}
{|f_{3\alpha}|\sqrt{f_{2\alpha}}},
\end{equation}
allowing the generalized Bloch factor to be written as
\begin{equation}
\beta_{\alpha}
=r_{\alpha}e^{ik}
=
\frac{|f_{1\alpha}|\sqrt{f_{4\alpha}}}
{|f_{3\alpha}|\sqrt{f_{2\alpha}}}\,e^{ik}.
\end{equation}

Therefore, while the individual solutions $\beta_{\alpha,1}(E)$ and $\beta_{\alpha,2}(E)$ retain an implicit dependence on the energy, the GBZ itself is energy-independent. The apparent reduction from three energy bands to two solutions for $\beta_{\alpha}$ reflects the fact that the characteristic equation defines a coupled relation between $\beta_{\alpha}$ and $E$, rather than yielding independent solutions for $\beta_{\alpha}$ alone.

\section{PBC solution}\label{App:PBCsolution}

To obtain Eqs. (\ref{eq:phase1})--(\ref{eq:phase3r}) of the main text, we use the characteristic equation of the Hamiltonian (\ref{Hbeta}) in the main text, which yields, 
\begin{equation}
    E^3 + p_{\alpha } E + q_{\alpha } = 0,
\end{equation}
where the momentum-dependent coefficients are
\begin{eqnarray}
&p_{\alpha } = 2f_{5\alpha }+f_{6\alpha },  \\
&q_{\alpha } =  2 i^{\frac{2-\operatorname{Sign}\left(f_{2\alpha }\right)-\operatorname{Sign}\left(f_{4\alpha }\right)}{2} }
\,|f_{5\alpha }|\sqrt{|f_{6\alpha }|}\cos(k).
\end{eqnarray}
The discriminant of this cubic equation is given by
\begin{equation}
\Delta_{\alpha } = 4p_{\alpha }^3 - 27q_{\alpha }^2.\label{Eq:app descr}  
\end{equation}
Given that $f_{j\alpha}$ ($j=1,2,...6$) are real parameters, the coefficient $p_{\alpha }$ is always real. However, $q_{\alpha }$ can be either purely real or purely imaginary, depending on the sign of
$f_{6\alpha }$, i.e.,
\begin{equation}
q_{\alpha } \in  \begin{cases}
 \mathbb{R}, & 
\text{if } 
f_{6\alpha } >0, \\[3mm]
 i\mathbb{R}, & 
\text{if } 
f_{6\alpha } <0.
\end{cases}  
\end{equation}
Applying Cardona's method, the three eigenvalues can be expressed as
\begin{align}\label{eq:eigenvalues}
E_{1,2\alpha } &= \frac{(1\pm i\sqrt{3})p_{\alpha }}{\sqrt[3]{12}F_{\alpha }^{1/3}} + \frac{(1\mp i\sqrt{3})F_{\alpha }^{1/3}}{\sqrt[3]{144}}, \nonumber \\
E_{3\alpha } &= -\frac{\sqrt[3]{2/3}p_{\alpha }}{F_{\alpha }^{1/3}} - \frac{F_{\alpha }^{1/3}}{\sqrt[3]{18}},
\end{align}
where $F_{\alpha } = -9 q_{\alpha } + \sqrt{-3\Delta_{\alpha }}.$ In the following, to systematically classify the spectrum, we analyze the two distinct regimes based on the character of $q_{\alpha}$.

\subsection{The real regime: $q_{\alpha} \in \mathbb{R}$ (where $f_{6\alpha} > 0$)}

In this regime, we denote the discriminant as $\Delta^r_{\alpha } = 4p_{\alpha }^3 - 27q_{\alpha }^2$. The superscript $r$ indicates that $q_\alpha \in \mathbb{R}$. Based on this discriminant, the feature of the eigenvalues can be determined as follows:
\begin{equation}
\begin{cases}
&\text{Three distinct real roots:} \\
& \Delta^r_{\alpha } > 0, \rightarrow
\quad p_{\alpha} > 0
\;\land\;
-\dfrac{2\sqrt{p_{\alpha}^{3}}}{3\sqrt{3}}
<
q_{\alpha}<\dfrac{2\sqrt{p_{\alpha}^{3}}}{3\sqrt{3}}, \\
&\rightarrow (f_{5\alpha}<0 \land f_{6\alpha}>-8f_{5\alpha}) \ | \ (f_{5\alpha}>0).
\\[3mm]
&\text{Multiple real roots (at least two equal):} \\
& \Delta^r_{\alpha} = 0, \rightarrow
q_{\alpha} = \pm  \dfrac{2 p_{\alpha}^{3/2}}{3\sqrt{3}}.\\
\\[3mm]
&\text{One real root and two complex conjugate roots:} \\
& \Delta^r_{\alpha} < 0,\rightarrow
\begin{cases}
p_{\alpha} < 0,  \rightarrow f_{\alpha 5}<0 \land -2 f_{\alpha 5}< f_{\alpha 6}< -8f_{\alpha 5} \\[2mm]
p_{\alpha} \ge 0 \land
|q_{\pm}| > \dfrac{2\sqrt{p_{\alpha}^{3}}}{3\sqrt{3}} \rightarrow 0<f_{6\alpha} <-2f_{5\alpha}.
\end{cases}\\ 
&\rightarrow \begin{cases}
f_{5\alpha} < 0 \ \text{and} \ -2f_{5\alpha} \le f_{6\alpha} < -8f_{5\alpha}, \\
f_{5\alpha} < 0 \ \text{and} \ 0 < f_{6\alpha} < -2f_{5\alpha}.
\end{cases}
\end{cases}
\end{equation}
Notably, under these constraints, there are strictly no purely imaginary solutions.
 
\subsection{The imaginary regime: $q_{\alpha} \in i\mathbb{R}$ (where $f_{6\alpha} < 0$)}

When $f_{6\alpha} < 0$, $q_{\alpha }$ is purely imaginary and can be written as $q_{\alpha } = i b_{\alpha }$ with $b_{\alpha } \in \mathbb{R}$. We look for purely imaginary solutions. Therefore, substituting $E = i v$ into the cubic equation gives a new cubic form as
\begin{equation}
i(v^3 + p_{\alpha }v + b_{\alpha }) = 0.
\end{equation}
The corresponding discriminant is $\Delta^i_{\alpha } = -4p_{\alpha }^3 - 27b_{\alpha }^2$, where the superscript $i$ denotes $q_\alpha \in i\mathbb{R}$. The root classification follows as\vahid{:}
\begin{equation}
\begin{cases}
& \text{Three distinct purely imaginary roots:} \\
&\Delta^i_{\alpha} > 0, \rightarrow p_{\alpha} < 0 , -\frac{2\sqrt{p^{3}_{\alpha}}}{3\sqrt{3}} < b_{\alpha} < \frac{2\sqrt{p_{\alpha}^{3}}}{3\sqrt{3}} \\
&\rightarrow (f_{5\alpha}>0 \land f_{6\alpha}<-8f_{5\alpha}) \ | \ (f_{5\alpha}<0).
  \\[2mm]
& \text{Multiple imaginary roots (at least two degenerate):} \\
&\Delta^i_{\alpha} = 0, \rightarrow  b_{\alpha} = \pm \frac{2 p_{\alpha}^{3/2}}{3\sqrt{3}}.
  \\[1mm]
  &\text{One purely imaginary and a $\pm$real pair roots:} \\
&\Delta^i_{\alpha} < 0, \rightarrow \begin{cases}
p_{\alpha} > 0,  \\[6pt]
\\
p_{\alpha} \le 0 , |b_{\alpha}| > \dfrac{2\sqrt{p_{\alpha}^{3}}}{3\sqrt{3}},
\end{cases}\\
&\rightarrow \begin{cases}
f_{5\alpha} > 0 \ \text{and} \ -8f_{5\alpha} \le f_{6\alpha} < -2f_{5\alpha}, \\
f_{5\alpha} > 0 \ \text{and} \  -2f_{5\alpha}<f_{6\alpha}<0.
\end{cases}
\end{cases}
\end{equation}

\section{Similarity transformation and edge-state energies of a general SSH$_3$ Hamiltonian}\label{App1}

Here we provide the details underlying the similarity transformation introduced in Sec.~\ref{s3} and derive the edge-state energies and localization criteria referenced there. 

Recall that the transformation matrix $S_\alpha$ is defined by
\begin{equation}
[S_{\alpha}]_{n+1,n+1} = r_{1\alpha}^{\frac{1}{2}\left(n-\left\lfloor \frac{n}{3}\right\rfloor\right)} \cdot r_{2\alpha}^{\frac{1}{2}\left\lfloor \frac{n}{3}\right\rfloor},
\end{equation}
where $[\cdot]_{nn}$ denotes the $n$-th diagonal element ($n=0,1,2,3,\ldots,3N-1$), $\lfloor \cdot \rfloor$ denotes the floor function, and $r_{1\alpha} = f_{1\alpha}/f_{3\alpha}$, $r_{2\alpha} = f_{2\alpha}/f_{4\alpha}$. Applying this transformation yields the mapped Hamiltonian:
\begin{eqnarray}\label{App:Eq:SimHamiltonian}
H'_{\alpha } &=& S_\alpha^{-1} H_{\alpha } S_\alpha\\
&=&\begin{pmatrix}
0 & T_{1\alpha } &  &  &  &  \\
T_{1\alpha } & 0 & T_{2\alpha } &  &  &  &  & O &  \\
 & T_{2\alpha } & 0 & T_{3\alpha } &  &  \\
 &  & T_{3\alpha } & 0 & T'_{1\alpha } &  \\
 &  &  & T'_{1\alpha } & 0 & T'_{2\alpha } &  &  &  &  \\
 &  &  &  & T'_{2\alpha } & 0 & T'_{3\alpha } &  &  &  \\
 &  &  &  &  & T'_{3\alpha } & 0 & \ddots &  \\
 & O &  &  &  &  & \ddots & 0 &  \\
 &  &  &  &  &  &  &  &
\end{pmatrix},\nonumber
\end{eqnarray}
where the component amplitudes are defined as,
\begin{equation}
\begin{aligned}
T_{1\alpha} &= \operatorname{Sign}(f_{3\alpha})\,
  i^{\frac{1-\operatorname{Sign}(f_{5\alpha})}{2}}
  \sqrt{|f_{5\alpha}|},\\
T_{2\alpha} &= \operatorname{Sign}(f_{5\alpha})\, T_{1\alpha},\\
T_{3\alpha} &= i^{\frac{2-\operatorname{Sign}(f_{2\alpha})
  -\operatorname{Sign}(f_{4\alpha})}{2}}
  \sqrt{|f_{6\alpha}|},\\
T_{1\alpha}' &= i^{\frac{1-\operatorname{Sign}(f_{5\alpha})}{2}
  (\operatorname{Sign}(f_{6\alpha})-1)} T_{1\alpha},\\
T_{2\alpha}' &= i^{\frac{1-\operatorname{Sign}(f_{5\alpha})}{2}
  (\operatorname{Sign}(f_{6\alpha})-1)} T_{2\alpha},\\
T_{3\alpha}' &= i^{\operatorname{Sign}(f_{6\alpha})-1} T_{3\alpha}.
\label{eq:app:HopPar}
\end{aligned}
\end{equation}
This transformation maps the original non-Hermitian Hamiltonian to a spectrally equivalent form that recovers the conventional bulk--boundary correspondence. It preserves the trimerized nearest-neighbor structure of the SSH$_3$-Rashba model while renormalizing the hopping amplitudes. The full Hamiltonian operates in a six-dimensional basis per unit cell, and the specific structure of the hopping and RSOC parameters allows the model to be block-diagonalized into two independent effective three-band subsystems with the effective hopping amplitudes defined above.

A closer examination of these transformed parameters reveals that each is either purely real or purely imaginary, depending on the signs of $f_{5\alpha}$ and $f_{6\alpha}$. Specifically, $T_{3\alpha}$ (and $T_{3\alpha}'$) is purely real when $f_{4\alpha}$ and $f_{2\alpha}$ share the same sign, i.e., $f_{6\alpha}=f_{4\alpha}f_{2\alpha}>0$, and purely imaginary when they have opposite signs ($f_{6\alpha}<0$). Likewise, $T_{1\alpha}$ and $T_{2\alpha}$ (along with $T_{1\alpha}'$ and $T_{2\alpha}'$) are purely real when $f_{5\alpha}>0$ and purely imaginary when $f_{5\alpha}<0$. This sign dependence determines the Hermitian/anti-Hermitian character of $H'_\alpha$ as already discussed in Sec.~\ref{s3}.

To derive the resulting edge-state energies, it is more transparent to first solve the problem for a generic three-sublattice SSH (SSH$_3$) chain with real-valued hoppings $t_1, t_2, t_3$, and only afterward identify them with their spin-resolved counterparts $T_{1\alpha}, T_{2\alpha}, T_{3\alpha}$.

Consider a one-dimensional $SSH_3$ chain comprising three sublattices ($A$, $B$, $C$) per unit cell n. The intra-cell hoppings are denoted by $t_1$ (between $A$ and $B$) and $t_2$ (between $B$ and $C$), while the inter-cell hopping is $t_3$ (between $C_n$ and $A_{n+1}$). The stationary Schrödinger equations for the wavefunction amplitudes in the n-th unit cell are
\begin{align}
E \, \psi^A_n &= t_1 \psi^B_n + t_3 \psi^C_{n-1}, \label{eq:schA} \\
E \, \psi^B_n &= t_1 \psi^A_n + t_2 \psi^C_n, \label{eq:schB} \\
E \, \psi^C_n &= t_2 \psi^B_n + t_3 \psi^A_{n+1}. \label{eq:schC}
\end{align}

\subsection{Exponential ansatz for general states}

We introduce an exponential ansatz for the amplitude distribution across unit cells as
\begin{equation}
(\psi_n^A, \psi_n^B, \psi_n^C)^T = (\psi^A, \psi^B, \psi^C)^T \, \xi^n,
\end{equation}
where $\xi$ is a complex propagation factor.  The modulus condition $|\xi | = 1$ indicates extended bulk states, whereas $|\xi| \neq 1$ corresponds to localized edge states. Substituting this ansatz into the Schrödinger equations yields,
\begin{align}\label{A5}
E \, \psi^A &= t_1 \psi^B + t_3 \psi^C \xi^{-1}, \nonumber \\
E \, \psi^B &= t_1 \psi^A + t_2 \psi^C,  \\
E \, \psi^C &= t_2 \psi^B + t_3 \psi^A \xi. \nonumber
\end{align}
This can be written as a homogeneous linear system:
\begin{equation}
\begin{pmatrix}
-E & t_1 & t_3 \, \xi^{-1} \\
t_1 & -E & t_2 \\
t_3 \, \xi & t_2 & -E
\end{pmatrix}
\begin{pmatrix} \psi^A \\ \psi^B \\ \psi^C \end{pmatrix} = 0.
\end{equation}
Nontrivial solutions exist only if the determinant vanishes:
\begin{equation}
\det
\begin{pmatrix}
-E & t_1 & t_3 \xi^{-1} \\
t_1 & -E & t_2 \\
t_3 \xi & t_2 & -E
\end{pmatrix} = 0.
\end{equation}
Expanding this determinant gives a general cubic equation for all states (both bulk and edge):
\begin{equation}
\label{eq:general_cubic}
E^3 - S E - P (\xi + \xi^{-1}) = 0,
\end{equation}
where
\begin{equation}
S = t_1^2 + t_2^2 + t_3^2, \qquad
P = t_1 t_2 t_3,
\end{equation}
or equivalently,
\begin{equation}\label{A10}
\xi + \xi^{-1} = X(E) \equiv \frac{E^3 - S E}{P}.
\end{equation}

\subsection{Edge states under OBC}

Under OBC, the wavefunction vanishes at the left boundary (cell $n=1$), i.e., $\psi_0^C$. Consequently, Eqs.~(\ref{A5}) reduce to
\begin{align}
E \, \psi^A &= t_1 \psi^B, \\
E \, \psi^B &= t_1 \psi^A + t_2 \psi^C, \\
E \, \psi^C &= t_2 \psi^B + t_3 \, \xi \, \psi^A.
\end{align}
Eliminating the sublattice amplitude $\psi_A,\psi_B,\psi_C$ gives the propagation factor as
\begin{equation}
\xi = - \frac{E t_3}{t_1 t_2}.
\end{equation}
Substituting this condition back into Eq.~(\ref{A10}) yields
\begin{equation}
- \frac{E t_3}{t_1 t_2} - \frac{t_1 t_2}{E t_3} = \frac{E^3 - S E}{P}.
\end{equation}
Simplifying this relation gets a biquadratic equation for the edge-state energies
\begin{equation}
E^4 - (S - t_3^2) E^2 + (t_1 t_2)^2 = 0.
\end{equation}
Defining $S' = S - t_3^2 = t_1^2 + t_2^2$, the solution for $E^2$ are 
\begin{eqnarray}
&E_{edge}^2 = \frac{S' \pm \sqrt{\Delta}}{2}=\frac{t_1^2 + t_2^2 \pm |t_1^2 - t_2^2|}{2},\nonumber \\
&\xi = - \frac{t_3}{t_2},\ ||\ \xi = - \frac{t_3}{t_1}.
\label{Eedge}
\end{eqnarray}
Here, the discriminant is non-negative: $\Delta=S'^2 - 4 (t_1 t_2)^2 = (t_1^2 - t_2^2)^2$.

In the present non-Hermitian system, the hoppings $t_{1,2}$ are replaced  by the effective parameters $T_{1,2\alpha}$ (as indicated in Eq.~(\ref{eq:app:HopPar})). Because $|T_{1\alpha}|=|T_{2\alpha}|$ the edge-state energies simplify to
\[
E_{edge,\alpha} = \pm T_{1\alpha}, \quad \xi_\alpha = - \frac{T_{3\alpha}}{T_{1\alpha}}.
\]
Since $T_{1\alpha}$ can be either real or imaginary, it is obvious that the corresponding edge-state energies  must also be purely real or purely imaginary.

\begin{figure}[t!]
    \centering     \includegraphics[width=8cm]{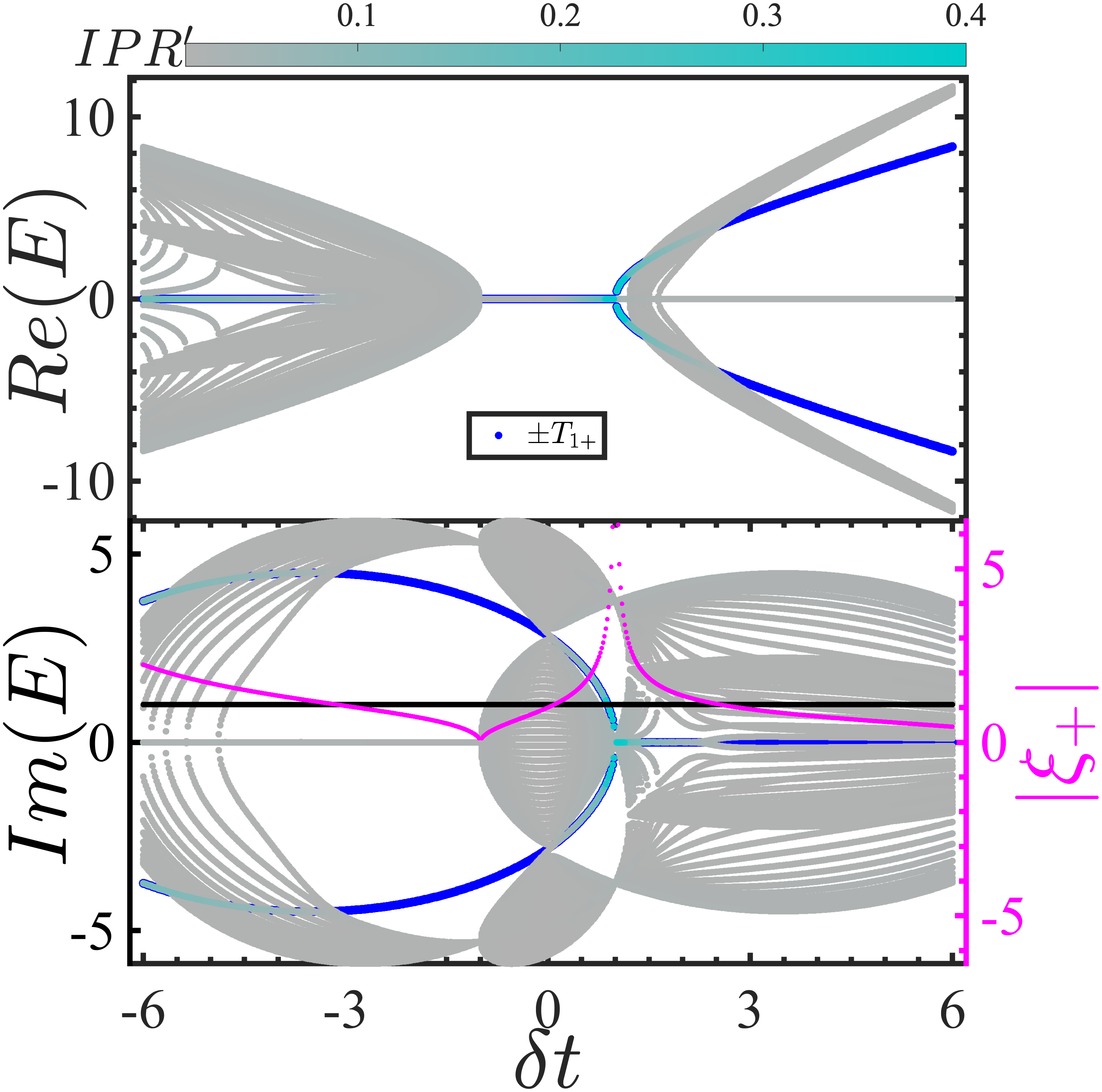}
    \caption{(Color online) Real and imaginary parts of the energy spectrum as a function of $\delta t$. The solid blue and magenta lines indicate the analytical edge-state energies $T_{1+}$ and localization decay parameter $|\xi_+|$ (right axis). The horizontal black line marks the localization threshold $|\xi_+|=1$. The same parameters are identical to those used in figure~\ref{fig:energy_levels_IPR_DIPR}(j).}
    \label{fig:appendix}
\end{figure}

Figure \ref{fig:appendix} illustrates both the real and imaginary energy spectra of spin-up sector, comparing the analytically derived edge energies (plotted as solid dark blue lines) to the in-gap edge states obtained through exact numerical diagonalization. The numerical states are quantified by IPR$'$. The strong alignment of the theoretical dark blue lines with the states exhibiting high IPR$'$ (cyan markers) confirms that these analytical solutions accurately capture the highly localized boundary modes. Notably, the analytical $T_{1 +}$ (blue line) matches the energy of the numerical edge states perfectly. 

Additionally, the lower panel of Fig. \ref{fig:appendix} plots the analytically obtained localization decay parameter $\xi$, as the magenta line. Exponentially localized edge states can exist only when $|\xi|>1$. The horizontal black reference line indicates this threshold ($|\xi|=1$). As $\delta t$ varies, the topological edge states (highlighted by cyan markers and tracked by the dark blue analytical lines) remain distinct and well-separated from the bulk spectrum, meaning their energies do not mix with those of the bulk states, only in the regions where the magenta curve stays above the black line. In contrast, when the magenta line crosses below the black line (e.g., in the region $\delta t \simeq [-5.5,-3.5])$, the edge-state energies merge with the bulk spectrum and their localization is destroyed, confirming our analytical results.

\section{Bernard–LeClair symmetry analysis and robustness of the edge states}\label{app: EdgeSymmetry}

To examine the topological origin and robustness of the edge states, we analyze the symmetries of the Hamiltonian within the Bernard--LeClair classification framework for non-Hermitian systems \cite{KawabataSym2019}.
This framework classifies systems based on
$C$ symmetry ($U_C H^T U_C^{-1} = \pm H$) and $Q$ symmetry 
($U_Q H^\dagger U_Q^{-1} =  H$). However, for our model, under the reality condition $H = H^*$, these two symmetry constraints reduce to a single transpose-type symmetry,
\begin{equation}
U H^T U^{-1} = \pm H,
\end{equation}
where their associated symmetry operators can obey different algebraic 
constraints (e.g., $U U^* = \pm 1$), which may still distinguish subclasses.

It is straightforward to show that each spin-resolved subsystem, $H_{\alpha}$ (for $\alpha=\pm$), exhibits two critical symmetries. The first is a $\mathcal{Q}_+$ symmetry defined by
\begin{eqnarray}
    \mathcal{Q}_+\, H_{\alpha }\, (\mathcal{Q}_+)^{-1} =H^{\dagger}_{\alpha}=H^{T}_{\alpha},
\end{eqnarray} 
where the operator is given by
\begin{eqnarray}\label{e3}
\mathcal{Q}_+ &=&
\begin{pmatrix}
 &   &        &   & 1 \\
 & O &        & 1 &   \\
 &   & \ddots &   &   \\
 & 1 &        & O &   \\
 1 &   &      &   &   
\end{pmatrix}_{3N}.
\end{eqnarray}
Furthermore, each subsystem possesses a $\mathcal{Q}_-$ symmetry, satisfying
\begin{eqnarray}\label{anti12}
    \mathcal{Q}_{-}\, H_{\alpha }\, (\mathcal{Q}_{-})^{-1}  = -H^{\dagger}_{\alpha }=-H^{T}_{\alpha},
\end{eqnarray}
where the corresponding operator is
\begin{eqnarray}\label{e6}
\mathcal{Q}_{-} &=&
\begin{pmatrix}
   &     &        &   & 1 \\
   & O   &        & -1&   \\
   &     & 1      &   &   \\
   & \ddots &     & O &   \\
   &     &        &   &   
\end{pmatrix}_{3N}.
\end{eqnarray}
Note that these operators satisfy $\mathcal{Q}_{-}^T\mathcal{Q}_{-}^{-1}=-1$, implying $\mathcal{Q}_{-}$ act as a particle-hole-like symmetry operator, and $\mathcal{Q}_{+}^T\mathcal{Q}_{-}^{-1}=\mathcal{Q}_{+}^\dagger\mathcal{Q}_{-}^{-1}=1$ representing pseudo-Hermiticity (or $TRS^\dagger$ since $H^\dagger=H^T$). Moreover, under PBCs, this symmetry preserve the  condition $\mathcal{Q}_{-}\, h_{\alpha}(k)\, (\mathcal{Q}_{-})^{-1}  = -h^{T}_{\alpha}(k+\pi)$ with the momentum shift $k\rightarrow k +\pi$~\cite{Jangjan2022}. 

To verify that the edge states are protected by $\mathcal{Q_+}$ and $\mathcal{Q_-}$ symmetries, we examine their stability by introducing perturbations of the form $H_{\alpha } \rightarrow H_{\alpha } + H_{\mathrm{pert}}$ and observing the spectral response. In what follows, we apply symmetry-preserving or symmetry-breaking perturbations as sublattice-dependent onsite potentials.

\begin{figure}[t!]
    \centering     \includegraphics[width=1\linewidth]{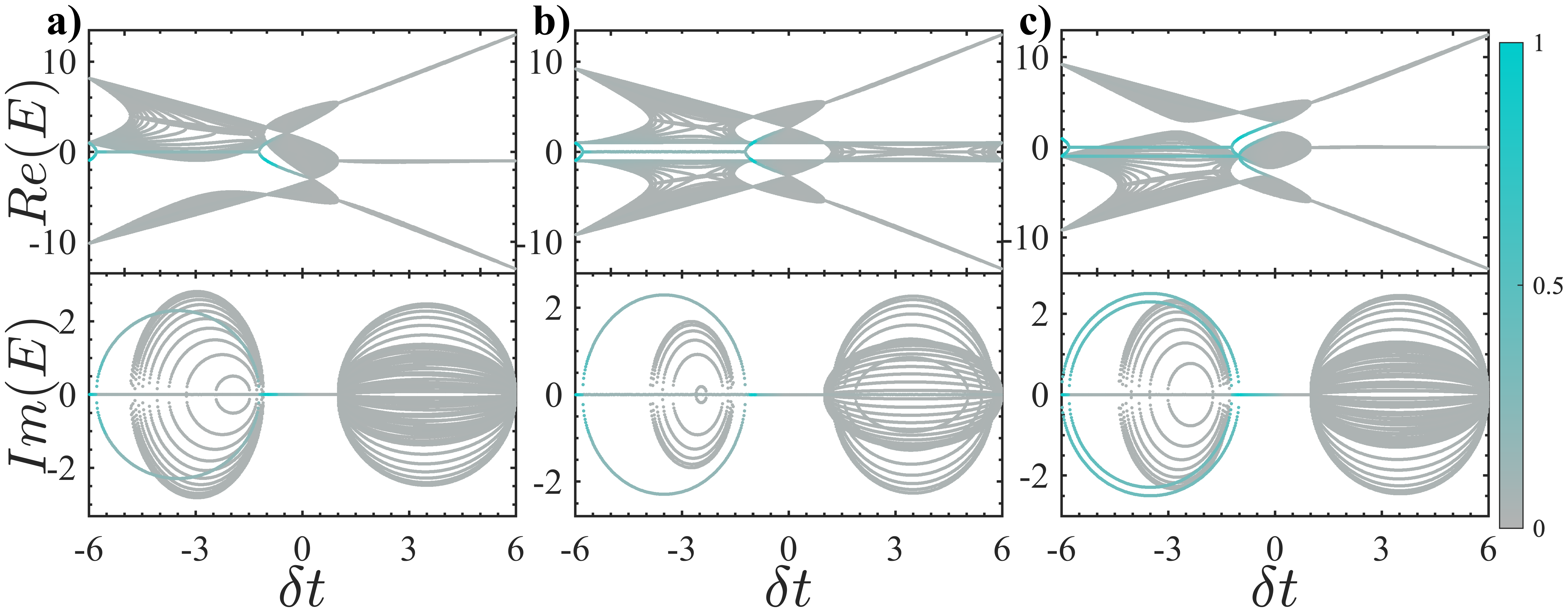}
    \caption{(Color online) Energy spectra of the  spin-down subsystem under OBCs as a function of $\delta t$ in the presence of three types of sublattice-dependent onsite perturbations. (a,b) Upon applying the first and second types of perturbations (which preserve $\mathcal{Q_+}$ and $\mathcal{Q_-}$ symmetries, respectively), the real parts of the edge-state energies remain pinned at zero, maintaining their degeneracy. (c) The third perturbation breaks both symmetries, resulting in the loss of degeneracy and causing the edge states to acquire finite real-energy components. Parameters are identical to those in Fig. \ref{fig:energy_levels_IPR_DIPR}(f) with the perturbation strength set to $v=1$.}
    \label{fig99}
\end{figure}

First, we consider a perturbation of the first kind as
\begin{equation}\label{eq:1Per}
H_{\mathrm{pert}} = v\,\sum_{n,\alpha}
\left(
-A_{n,\alpha}^\dagger A_{n,\alpha}
+B_{n,\alpha}^\dagger B_{n,\alpha}
- C_{n,\alpha}^\dagger C_{n,\alpha}
\right),
\end{equation}
or a second kind as
\begin{equation}\label{eq: 2Per}
H_{\mathrm{pert}} = v\,\sum_{n,\alpha} (-1)^n
\left(
- A_{n,\alpha}^\dagger A_{n,\alpha}
+ B_{n,\alpha}^\dagger B_{n,\alpha}
- C_{n,\alpha}^\dagger C_{n,\alpha}
\right),
\end{equation}
where $v$ is the strength of the perturbation. The first (second) kind of perturbation preserves $\mathcal{Q_+}$ ($\mathcal{Q_-}$) symmetry. Figures~\ref{fig99}(a) or \ref{fig99}(b) show the energy spectra under OBCs of the system in the presence of the first or the second perturbation, respectively. In both panels, it is obvious that although the bulk spectrum is modified and exceptional points may shift, the real parts of the edge-state energies remain pinned at zero, preserving their degeneracy. This demonstrates that these edge modes are topologically protected by the $\mathcal{Q_+}$ or $\mathcal{Q_-}$ symmetry rather than by a fine-tuned choice of parameters.

In contrast, we apply a third kind of perturbation 
\begin{equation}
H_{\mathrm{pert}} = v\,\sum_{n,\alpha}
\left(
 A_{n,\alpha}^\dagger A_{n,\alpha}
- B_{n,\alpha}^\dagger B_{n,\alpha}
- C_{n,\alpha}^\dagger C_{n,\alpha}
\right),
\end{equation}
to the unperturbed Hamiltonian. This perturbation explicitly breaks both the $\mathcal{Q_+}$ and $\mathcal{Q_-}$ symmetries. The corresponding energy spectrum is presented in Fig.~\ref{fig99}(c). In this case, the zero-energy degeneracy is lifted, and the edge states immediately acquire finite real-energy components, hybridizing with nearby bulk states.

This robustness against symmetry-preserving perturbations, combined with the edge-only $\mathcal{APT}$ symmetry breaking discussed in Secs.~\ref{EdgeSolve}, demonstrates that the spin-resolved SSH$_3$-Rashba model realizes edge-selective symmetry preservation. Specifically, different symmetry classes ($\mathcal{PT}$ and $\mathcal{APT}$) govern bulk and edge simultaneously, with the topologically edge states being protected independently by the underlying $\mathcal{Q}_+$ and $\mathcal{Q}_-$ symmetries.

\bibliography{References} 

\end{document}